\documentclass[twocolumn,eqsecnum,showpacs,showkeys,amsmath,amssymb,nofootinbib,superscriptaddress,floatfix]{revtex4-1}
\usepackage{graphicx}
\usepackage{subfigure}
\usepackage{bm}

\makeatletter
\newcommand\erfc{\mathop{\operator@font erfc}\nolimits}
\def\slashchar#1{\setbox0=\hbox{$#1$}
   \dimen0=\wd0 \setbox1=\hbox{/} \dimen1=\wd1
   \ifdim\dimen0>\dimen1 \rlap{\hbox to \dimen0{\hfil/\hfil}} #1
   \else  \rlap{\hbox to \dimen1{\hfil$#1$\hfil}} / \fi}

\makeatother

\begin{document}
\title{Collective flow in p-Pb and d-Pb collisions at TeV energies}
\author{Piotr Bo\.zek}
\email{Piotr.Bozek@ifj.edu.pl}
\affiliation{The H. Niewodnicza\'nski Institute of Nuclear Physics,
PL-31342 Krak\'ow, Poland} \affiliation{
Institute of Physics, Rzesz\'ow University, 
PL-35959 Rzesz\'ow, Poland}
\date{\today}

\begin{abstract}
We apply the hydrodynamic model to the dynamics of  matter created in p-Pb collisions at
$\sqrt{s_{NN}}=4.4$TeV  and d-Pb collisions at $\sqrt{s_{NN}}=3.11$TeV.
The fluctuating initial conditions are calculated in the Glauber Monte-Carlo model for several
centrality classes. The  expansion is performed event by event  in $3+1$-dimensional viscous 
hydrodynamics.  Noticeable elliptic and triangular flows appear 
 in the distributions of produced particles.
\end{abstract}

\pacs{25.40.Ep, 25.45.De, 13.85Hd, 25.75.Ld}

\keywords{hadron-nucleus
 collisions, viscous  hydrodynamic model, collective flow}

\maketitle

\section{Introduction}

The large multiplicity of particles emitted from the small interaction region in relativistic heavy-ion collisions implies that a  fireball  of very dense matter is formed. Experiments at  
 the BNL Relativistic Heavy Ion Collider (RHIC)  
and the CERN Large Hadron Collider (LHC)
\cite{Arsene:2004fa,*Back:2004je,*Adams:2005dq,*Adcox:2004mh,Aamodt:2010pa,*Aamodt:2011mr,*Aad:2010bu,*Chatrchyan:2011sx} have demonstrated 
the appearance of a collective flow in the expanding fireball.
The  physical picture is expected to be different in the interaction of small systems,
proton-proton, proton-nucleus or deuteron-nucleus.  At RHIC energies the density of matter created in
 d-Au interactions is small and does not cause jet quenching \cite{Arsene:2004fa}. d-Au and p-p 
interactions are treated as a baseline reference to evidence new effects in nucleus-nucleus collisions,
beyond a simple superposition of nucleon-nucleon (NN) collisions.
With the advent of proton-proton collisions at several TeV center of mass (c.m.)
 energies at the LHC, it has been 
suggested that some degree of collective expansion appears in high multiplicity
p-p events
 \cite{Luzum:2009sb,d'Enterria:2010hd,*Bozek:2009dt,*CasalderreySolana:2009uk,*Avsar:2010rf,Khachatryan:2010gv,Aamodt:2011kd,*Bozek:2010pb,*Werner:2010ss}. 
However, 
no direct experimental evidence exist for such  a collective expansion in p-p interactions. 

At the LHC, p-Pb collisions can be  studied in the future, experiments with 
d-Pb or other asymmetric systems 
are also possible, but with additional technical difficulties \cite{Salgado:2011wc}.
Estimates of the hadron production in p-Pb interactions  at TeV energies take into account nuclear 
effects on the parton distribution functions, saturation effects, 
but do not assume the formation of a hot medium 
\cite{Dumitru:2011wq,*Barnafoldi:2011px,*Kharzeev:2004if,*QuirogaArias:2010wh,*Rezaeian:2009it,*Salgado:2011pf,Salgado:2011wc}.  Experiments with p-Pb beams should provide an input for 
models used in heavy-ion collisions for the calculation of dense  medium effects on hard-probes.

The expected multiplicity and  size of the interaction region in central p-Pb and d-Pb
collisions at TeV energies are similar as in peripheral ($60-80$\% centrality) 
Pb-Pb  collisions  at $\sqrt{s_{NN}}=2.76$TeV \cite{Aamodt:2010cz}. This raises the possibility that hot
and dense matter is formed in such collisions. For  strongly interacting matter, the assumption 
of local equilibrium is a good approximation and relativistic  hydrodynamics 
can be used to follow the evolution of the system \cite{Kolb:2003dz,*Huovinen:2006jp,*Florkowski:2010zz}. 
Quantitative predictions for the elliptic flow have to  account for finite deviations 
from local equilibrium in the rapidly expanding fluid 
\cite{Luzum:2008cw,*Chaudhuri:2006jd,*Huovinen:2008te,*Song:2007fn,*Dusling:2007gi,Bozek:2009dw,Schenke:2010rr}. 

In order to test the assumption of the formation of a dense fluid in p-Pb and d-Pb interactions
 and to estimate possible effects of its collective expansion, we apply the viscous hydrodynamic
 model to calculate the spectra of emitted particles. The goal of this study is to have
 a quantitative prediction of the 
elliptic and triangular flows and of the transverse momentum spectra
for comparison with future experiments. The dynamically evolved density  of the fireball from  
hydrodynamic simulations can be used in the calculations of the parton energy loss in such 
small systems. 

The task requires the use of the most sophisticated version of the hydrodynamical model:
 event by event $3+1$-dimensional
($3+1$-D) viscous hydrodynamics. While
a good description of many collective phenomena in heavy-ion collisions 
can be obtained in the perfect fluid hydrodynamics 
in $2+1$-D \cite{Kolb:2003dz,Teaney:2000cw,*Huovinen:2003fa,*Broniowski:2008vp,*Akkelin:2008eh} 
or $3+1$-D \cite{Hirano:2002ds,*Hama:2005dz,*Nonaka:2007nn,*Bozek:2009ty},
to calculate the azimuthally asymmetric flow in  small systems such as p-Pb or d-Pb collisions
 one has to use 
viscous hydrodynamics. In collisions of symmetric nuclei $2+1$-D boost-invariant
 viscous hydrodynamics is routinely being applied for observables at central rapidities
 \cite{Luzum:2008cw,Bozek:2009dw}. 
In p-Pb or d-Pb interactions the energy density and the 
final particle distributions depend strongly
on rapidity. This forces the use of $3+1$-D hydrodynamics to obtain realistic 
particles spectra at different rapidities. Only recently $3+1$-D viscous hydrodynamic 
simulations became available \cite{Schenke:2010rr,Bozek:2011ua}. 
In proton or deuteron interaction with a nucleus the shape of the interaction region 
fluctuates widely from event to event. Unlike in interactions of heavy ions, using the average density 
is not a reliable approximation. Event by event $3+1$-D perfect fluid
 hydrodynamics 
is used by several groups 
\cite{Andrade:2006yh,*Werner:2009fa,*Petersen:2010cw}. 
The inclusion 
of event by event  fluctuations is important in the description 
of the initial eccentricity and triangularly of the fireball 
\cite{Alver:2008zz,Andrade:2006yh,Alver:2010gr,Alver:2010dn,Schenke:2010rr,Qiu:2011fi}.  
Only one group is using an event by event $3+1$-D viscous hydrodynamic 
code for heavy-ion collisions \cite{Schenke:2010rr,Schenke:2011bn,*Schenke:2011tv}.

As the size and the 
life-time of the system decrease the hydrodynamic model becomes less justified.  
A sizable elliptic flow is observed in peripheral Pb-Pb collisions at the LHC, which proves that 
substantial rescattering occurs in the evolution of the fireball. By itself it does not prove that 
the hydrodynamic regime is applicable in such collisions, as some  elliptic flow
 can be generated through collisions in the dilute limit. Few hydrodynamic calculations are applied 
also to peripheral Pb-Pb collisions at $\sqrt{s}=2.76$TeV \cite{Luzum:2009sb,Bozek:2011wa,Shen:2011eg}
with results compatible with experimental observations. Nevertheless,
it must be noted that as the impact parameter increases,
uncertainties of the hydrodynamic model become more important; fluctuations modify substantially 
the initial eccentricity, the relative role of the hadronic corona in 
the evolution of the system increases. In the present calculation the last issue is 
partly taken into account through an increase of the shear viscosity to entropy ratio 
at lower temperatures.  For d-Au collisions at $\sqrt{s_{NN}}=200$GeV the hydrodynamic model is expected 
to break down, as indicated by the absence of jet quenching. However, there are no published experimental
 results concerning directly the elliptic flow in d-Au collisions or estimates from hydrodynamic models
at RHIC energies.

Below we present results from  event by event viscous hydrodynamic simulations for
p-Pb and d-Pb collisions at $\sqrt{s_{NN}}=4.4$ and $3.11$TeV respectively. We use Glauber 
Monte-Carlo model initial conditions for the hydrodynamic evolution. 
We calculate particle spectra, charged particle pseudorapidity distributions, 
elliptic and triangular flow coefficients as function of pseudorapidity 
and transverse momentum.

\section{Size and shape of the initial fireball}

\label{sec:ini}

The number of particles produced in a p-Pb or d-Pb interaction 
can be estimated from $N_{part}$ the number of participant (wounded) 
nucleons in collision.
The Glauber Monte-Carlo model generates a distribution 
of events with different source sizes (number of participant nucleons)
 and different shapes
(distribution of participant nucleons in the transverse plane). The binary
 collision contribution is expected to be numerically small. Moreover  the
number of  binary collisions is roughly $N_{part}-1(2)$. The
presence of a term depending on the number of binary collisions
cannot be separated from the  functional dependence on $N_{part}$.
The number of participant nucleons in the Glauber model 
depends on the NN cross section. 

\begin{table}
\begin{tabular}{|c|c|c|c|}
\hline
system & $\sqrt{s_{NN}}$ & $\sigma_{NN}$ & $\frac{dN}{d\eta_{PS}}$ \\
& TeV & mb & at $\eta_{ps}=0$  \\
\hline
\hline
p-Pb & $4.4$ & $66.4$ & $ 50\pm 5$ \\
& 8.8 & $73.4$ & $65 \pm 9$ \\
\hline
d-Pb & $3.11$ & $63$ & $80 \pm 5$ \\
& $6.22$ & $69.8$ & $95 \pm 10$ \\
\hline
\end{tabular}
\caption{NN cross section (third column) and expected density of charged 
particles at mid-rapidity in the 
NN c.m. (fourth column) for different systems and energies
in central collisions.}
\label{tab:data}
\end{table}

p-Pb interactions at the 
LHC are planed at the c.m. energy in the NN system starting
 at $\sqrt{s_{NN}}=4.4T$TeV. This corresponds to proton and Pb  momenta of 
$3.5$TeV and $208\times 1.38$TeV, attainable with the present magnetic field 
configurations in the accelerator \cite{Salgado:2011wc}. For deuteron 
beams it gives the energy $\sqrt{s_{NN}}=3.11$TeV. The maximal available 
NN c.m. energy at the LHC is $8.8$ and $6.22$TeV for p-Pb and d-Pb 
interactions respectively. 
For collisions of beams with different energies per nucleon, the NN c.m.
  reference frame is shifted in rapidity with respect to the 
laboratory frame. The shift is   $y_{sh}=0.46$ and $0.12$ for p-Pb and d-Pb 
interactions. All the calculations in the hydrodynamic model are  made in 
the NN c.m. frame. For the final emitted particles a boost is made to 
the laboratory frame to get spectra around mid-rapidity or pseudorapidity
 distributions.

The NN cross section at different energies can be obtained from an 
interpolation of values at  $200$GeV $2.76$TeV and $7$TeV   
\cite{Loizides:2011ys,Aad:2011eu}
($\sigma_{NN}=42$, $62$, and $71$mb respectively)
using a formula of the form
$\sigma_{NN}\propto a +b \ln(\sqrt{s_{NN}}) +c \ln^2(\sqrt{s_{NN}})$.
The resulting NN cross sections from the Table \ref{tab:data}
are used in our Glauber model calculation. We take a Wood-Saxon profile
for the Pb nuclear density 
\begin{equation}
\rho(x,y,z)=\frac{\rho_0}{1+\exp\left((\sqrt{x^2+y^2+z^2}-R_A)/a\right)} \ ,
\end{equation}
with   $\rho_0=0.17 \mbox{fm}^{-3}$, $R_A=6.55$fm and
$a=0.45$fm, and an excluded distance for nucleons of $0.4$fm,
 for the deuteron we use  the Hulthen distribution \cite{Broniowski:2007nz}.

\begin{figure}
\includegraphics[width=.35\textwidth]{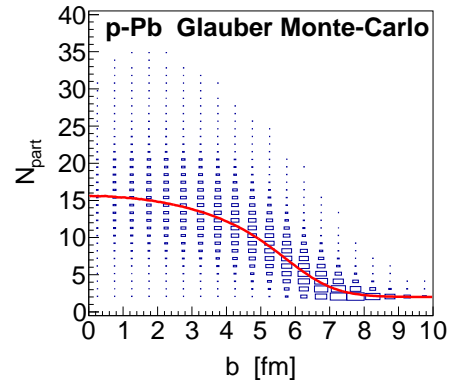}
\caption{(Color online) The distribution of participant nucleons at different impact
 parameters (boxes) and the average number of nucleons as function of the impact parameter 
(solid line) for p-Pb interactions.}
\label{fig:nwevp}
\end{figure}


\begin{figure}
\includegraphics[width=.45\textwidth]{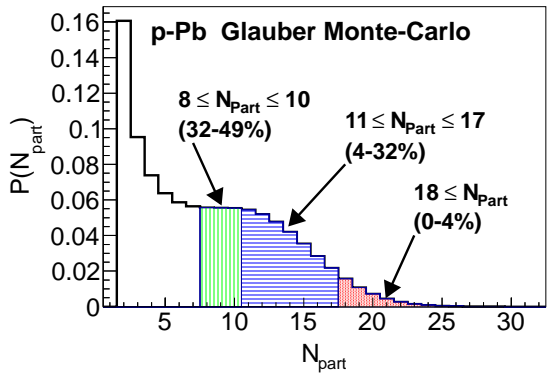}
\caption{(Color online) The probability distribution of participant nucleons in  p-Pb interactions. 
The three centrality classes considered in the simulations are defined by cuts in the number of participant nucleons.}
\label{fig:nwpprob}
\end{figure}

\begin{figure}
\includegraphics[width=.45\textwidth]{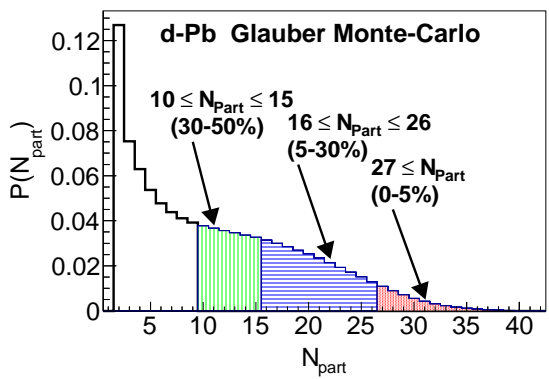}
\caption{(Color online) Same as Fig. \ref{fig:nwpprob} but for d-Pb interactions.}
\label{fig:nwdprob}
\end{figure}

Events at a given impact parameter are generated using 
the GLISSANDO code for the Glauber model \cite{Broniowski:2007nz}.
The distribution of participant nucleons at different
 impact parameters is shown in Fig. \ref{fig:nwevp} for p-Pb
 interactions at $4.4$TeV. We notice that the number of participant 
nucleons fluctuates strongly at a fixed impact parameter. 
The number of participant nucleons can
 be significantly above the average value (solid line in Fig.
 \ref{fig:nwevp}). Defining the most central collisions as a interval in 
the impact parameter is incorrect. The few percents 
most central events in terms of
 the number of participant nucleons ($N_{part}>18$) have 
a participant multiplicity larger than the average $N_{part}$ at zero impact parameter.
The picture is very similar for d-Pb collisions.
In the experiment the centrality classes are defined by the track multiplicity,
closely correlated with the number of participants in the model.
In heavy-ion collisions the number of participants is correlated with the 
impact parameter. In p-Pb or d-Pb interaction it is preferable to define
the centrality classes for events 
using directly cuts in $N_{part}$. Figs. \ref{fig:nwpprob} and \ref{fig:nwdprob}
show the probability density for events of a given $N_{part}$ for the two 
systems considered. For p-Pb events, 
we use three centrality classes  defined as
$18\le N_{part}$, $11\le N_{part}\le 17$, and $8\le N_{part}\le 10$ corresponding
 to centrality bins $0-4$\%, $4-32$\% and $32-49$\% out of  all the inelastic
 events  ($N_{part}\ge 2$). The unusual numbers for the  centrality  percentiles are 
 fixed by the discrete
variable $N_{part}$.  For the d-Pb interactions, we choose 
$27\le N_{part}$, $16\le N_{part}\le 26$, and $10\le N_{part}\le 15$ corresponding
 to centrality bins $0-5$\%, $5-30$\% and $30-50$\%.

The charged particle density at central pseudorapidity can be 
estimated from
 the multiplicity observed at a similar energy and for a similar number
of
participant nucleons  measured in peripheral Pb-Pb collisions at 
the LHC \cite{Aamodt:2010cz}, interpolating the measured values 
of ${dN}/d\eta_{PS}\ / \ <N_{part}/2>$ at   centralities $60-70$\% and $70-80$\%
to the average number of participant nucleons 
$\langle N_{part}\rangle $ corresponding to the most central 
bins considered in p-Pb
 and d-Pb collisions. The   energy dependence of $dN/\eta_{PS}$ 
is $s^{0.11}$ for p-p and $s^{0.15}$ for nucleus-nucleus collisions 
\cite{Toia:2011rt}. We take 
  $s^{0.13}$ to extrapolate from $\sqrt{s_{NN}}=2.76$TeV. The estimated values of 
the charged particle density at midrapidity are quoted in Table \ref{tab:data},
the uncertainty comes from the uncertainty in the 
measurements  \cite{Aamodt:2010cz} and in the value of the exponent in the
 energy dependence. 

\begin{figure}
\includegraphics[width=.38\textwidth]{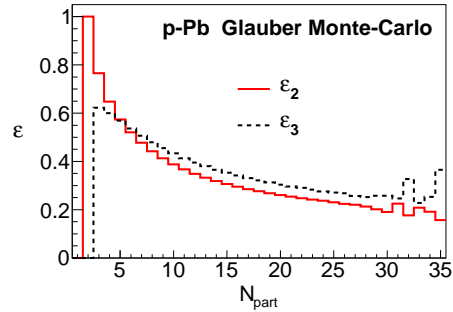}
\caption{(Color online) Eccentricity (solid line) and triangularity
 (dashed line) 
 in p-Pb interactions as function of the number of participant nucleons.}
\label{fig:e2pPb}
\end{figure}

\begin{figure}
\includegraphics[width=.38\textwidth]{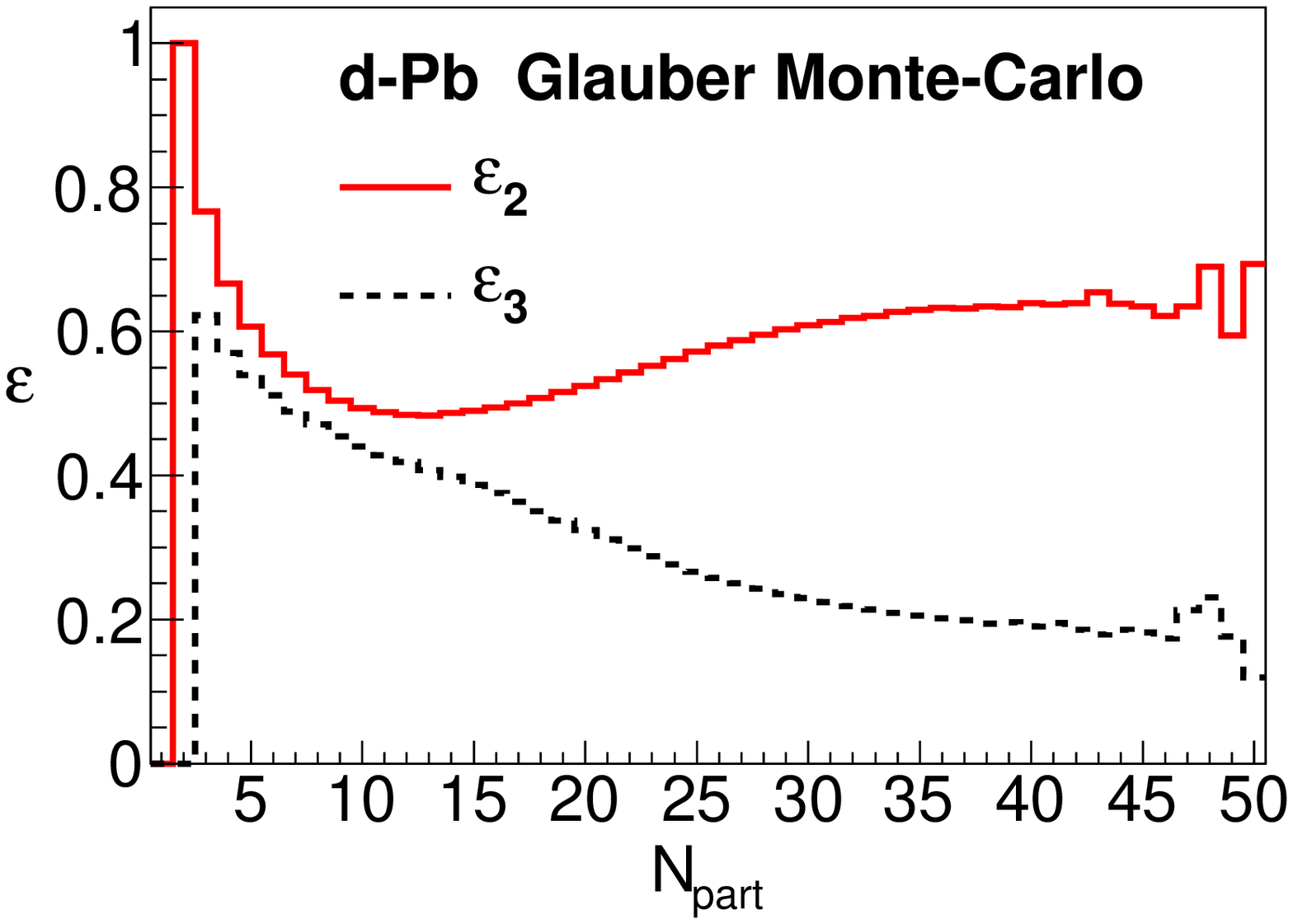}
\caption{(Color online) Same as Fig. \ref{fig:e2pPb} but for d-Pb interactions.}
\label{fig:e2dPb}
\end{figure}

\begin{figure}
\includegraphics[width=.3\textwidth]{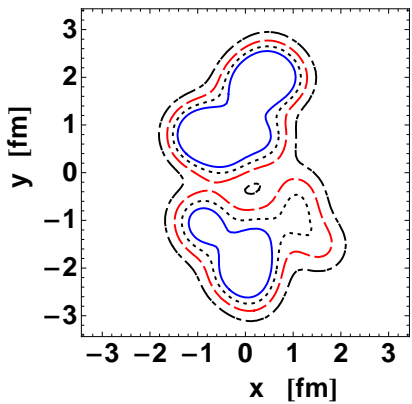}
\caption{(Color online) Contour plot $s(x,y,\eta_\parallel=0)$ 
of the initial entropy density in a d-Pb collision with $N_{part}=24$. }
\label{fig:inidxy}
\end{figure}

The azimuthally asymmetric collective flow is driven by the 
asymmetry of the initial fireball. The initial eccentricity in events
with $N_{part}$ participant nucleons
\begin{equation}
\epsilon_2=\frac{\langle \sum_{i=1}^{N_{part}}r^2_i \cos\left(2(\phi_i-\psi_2)\right) \rangle}
{< \sum_{i=1}^{N_{part}}r^2_i >}
\end{equation}
is calculated in each event with respect to the eccentricity 
angle $\psi_2$ maximizing $\epsilon_2$, the sum runs over all participant nucleons at positions $r_i$, $\phi_i$, $\langle \dots \rangle$ denotes averaging over events.
In a similar way the triangularity is 
\begin{equation}
\epsilon_3=\frac{\langle \sum_{i=1}^{N_{part}}r^3_i \cos\left(3(\phi_i-\psi_3)\right)\rangle}
{\langle \sum_{i=1}^{N_{part}}r^3_i \rangle}  \ ,
\end{equation}
and is calculated with respect the triangularity axis $\psi_3$ in each event 
\cite{Alver:2008zz,Alver:2010gr}. In Figs. \ref{fig:e2pPb} and \ref{fig:e2dPb}
we plot the eccentricity and the triangularity as function of the number 
of participant nucleons in the fireball. In proton induced interactions, the 
eccentricity and the triangularity of the source are similar and decrease for 
central collisions. It is different for d-Pd collisions, the eccentricity is 
larger than the triangularity and increases for central events. The eccentricity in d-Pb interactions
is caused by the asymmetric configuration of the two nucleons in the deuteron. Configurations with a large separation 
of the deuteron proton and neutron in the transverse plane have a large eccentricity and 
usually lead 
to a large number of participant nucleons  in the Pb nucleus. This effect 
causes the increase of the eccentricity for the most central collisions in Fig. \ref{fig:e2dPb}.
We note that the eccentricity in Glauber models can be modified by 
correlation effects 
\cite{Broniowski:2007nz,Broniowski:2010jd,*Rybczynski:2011wv}


We assume that the 
initial entropy density in the fireball  is proportional
 to the number of participant nucleons. The 
density in the transverse plane $x$, 
$y$ is the sum of  contributions from  participant nucleons at positions
 $x_i$, $y_i$ from the Pb 
nucleus $N_{-}(x,y)$ and from the proton $N_+(x,y)$ 
(or from the proton and/or the neutron in the deuteron)
\begin{equation}
N_{\pm}(x,y)=s_0 \sum_i \frac{1}{2\pi \sigma_w^2} 
\exp \left(-\frac{(x-x_i)^2+(y-y_i)^2}{2 \sigma_w^2}\right) \ .
\label{eq:inis}
\end{equation}
The  contribution from each nucleon is a Gaussian of width $\sigma_w=0.4$fm.
The final results show some dependence on the chosen value of $\sigma_w$. Using a smaller width of 
$0.3$fm/c we notice  an  increase by $\simeq10\%$ of the integrated elliptic and triangular 
flow for p-Pb collisions. A similar effect has been  observed in Ref. \cite{Schenke:2011bn}.
The  parameter $s_0$ is fixed to reproduce the final multiplicity after 
the hydrodynamic evolution. The distribution in space-time rapidity 
$\eta_\parallel$ 
is  asymmetric
\begin{equation}
s(x,y,\eta_\parallel)=f_-(\eta_\parallel) N_-(x,y)+f_+(\eta_\parallel)N_+(x,y) \ , 
\label{eq:sdens} 
\end{equation}
the profiles  $f_{\pm}(\eta_\parallel)$ are 
of the form 
\begin{equation}
f_{\pm}(\eta_\parallel)=\left(1\pm \frac{\eta_\parallel}{y_{beam}}\right)f(\eta_\parallel) \ ,
\label{eq:asy}
\end{equation}
$y_{beam}$ is the beam rapidity in the NN c.m. frame.

The asymmetric emission in the forward (backward) rapidity hemisphere 
from forward  (backward) going nucleons 
can be observed in the distribution of charged particles in 
d-Au collisions at RHIC \cite{Bialas:2004su}. The  distribution of the form 
(\ref{eq:sdens}) has been used as the initial condition for the hydrodynamic 
evolution in modeling  Au-Au collisions at RHIC yielding a satisfactory 
description of the directed flow \cite{Bozek:2010bi}.
The parameters of the longitudinal profile
\begin{equation}
f(\eta_\parallel)=\exp\left(-\frac{(|\eta_\parallel|-\eta_0)^2}
{2\sigma_\eta^2}\theta(|\eta_\parallel|-\eta_0)
\right) \ ,
\label{eq:etaprofile}
\end{equation}
the plateau width $2\eta_0$ and the width of the Gaussian tails $\sigma_\eta$
are adjusted as initial conditions for  $3+1$-D viscous hydrodynamic 
calculations to reproduce the charged particle pseudorapidity distributions
in Au-Au collisions at $200$GeV ($\eta_0=1.5$, $\sigma_\eta =1.4$
 \cite{Bozek:2011ua}) and Pb-Pb collisions at $2.76$TeV \cite{Toia:2011rt} 
($\eta_0=2.3$, $\sigma_\eta =1.4$). For the present calculation we take 
$\sigma_\eta=1.4$ and $\eta_0=2.35$, $2.4$ for interactions at $3.11$ and 
$4.4$TeV respectively. An example of the initial entropy density in a d-Pb 
interaction event in shown in Fig. \ref{fig:inidxy}. Typically we observe 
strongly deformed lumpy initial states. The elongated shape of the source results from the configuration of the 
nucleons in the deuteron while hitting the larger nucleus. This configuration is 
more important for the resulting eccentricity and the total number of participant nucleons than the impact parameter
 (as long as the deuteron hits the core of the Pb nucleus).

\section{Viscous hydrodynamics}

\label{sec:visc}

We use the second order relativistic viscous hydrodynamics to evolve the initial
 energy density in each event \cite{IS}.  
The initial entropy density is generated 
in the Glauber Monte-Carlo procedure described in the previous section.
The viscous hydrodynamics incorporates deviations from 
local equilibrium in terms of the shear and bulk viscosities, 
at zero baryon density heat conductivity can be neglected.
 These corrections $\pi^{\mu\nu}$ and $\Pi$ 
to the energy momentum tensor $T^{\mu\nu}$ are evolved
 dynamically 
\begin{equation}
\Delta^{\mu \alpha} \Delta^{\nu \beta} u^\gamma \partial_\gamma 
\pi_{\alpha\beta}=\frac{2\eta \sigma^{\mu\nu}-\pi^{\mu\nu}}{\tau_{\pi}}
-\frac{4}{3}\pi^{\mu\nu}\partial_\alpha u^\alpha
\label{eq:pidyn}
\end{equation}
and
\begin{equation}
 u^\gamma \partial_\gamma \Pi=
\frac{-\zeta \partial_\gamma u^\gamma-\Pi}{\tau_{\Pi}}
-\frac{4}{3}\Pi\partial_\alpha u^\alpha  \ .
\label{eq:budyn}
 \end{equation}
$\Delta^{\mu\nu}=g^{\mu\nu}-u^\mu u^\nu$, 
\begin{equation}
\sigma_{\mu\nu}=\frac{1}{2}\left( \nabla_\mu  u_\nu
+\nabla_\mu   u_\nu    -\frac{2}{3}\Delta_{\mu    \nu  }\partial_\alpha
   u^\alpha\right)\ ,
\end{equation}
and
$\nabla^\mu=\Delta^{\mu\nu} \partial_\nu$.
The hydrodynamic equations
\begin{equation}
\partial_\mu T^{\mu\nu}=0
\end{equation}
are solved numerically in the proper time $\tau=\sqrt{t^2-z^2}$ on a grid in the transverse coordinates $x$, $y$ and the space-time rapidity $\eta_\parallel$, starting from $\tau_0=0.6$fm/c. We use $s_0=0.72$GeV$^3$ in (\ref{eq:inis}) 
for both p-Pb and
 d-Pb collisions, which gives the expected final multiplicities. 
We take for the  relaxation time $\tau_\pi=\frac{3\eta}{T s}$, and assume
$\tau_\Pi=\tau_\pi$. The initial fluid velocity $u^\mu$ is 
taken as the Bjorken flow, 
the initial stress corrections from shear viscosity correspond to the Navier-Stokes formula,
 while the initial bulk viscosity corrections are zero $\Pi(\tau_0)=0$.
The details of the solution in $2+1$-D and $3+1$-D are given 
in \cite{Bozek:2009dw,Bozek:2011ua}.

The shear viscosity to entropy ratio in our calculation 
is not constant. It takes the value $\eta/s=0.08$ in the plasma phase 
and increases in the hadronic phase \cite{Bozek:2011ua}
\begin{equation}
\frac{\eta}{s}(T)=\frac{\eta_{HG}}{s} f_{HG}(T)+(1-f_{HG}(T))\frac{\eta_{QGP}}{s}
\label{eq:etas}
\end{equation}
with $\eta_{HG}/s=0.5$, $\eta_{QGP}/s=0.08$, and  $f_{HG}(T)=1/\left(\exp\left((T-T_{HG})/\Delta T\right)+1\right)$, where $T_{HG}=130$MeV, $\Delta T=30$MeV. The
bulk viscosity is nonzero  in the hadronic phase
\begin{equation}
\frac{\zeta}{s}(T)=\frac{\zeta_{HG}}{s} f_{\zeta}(T)
\label{eq:zetas}
\end{equation}
with $\zeta_{HG}/s=0.04$ and  $f_{\zeta}(T)=1/\left(\exp\left((T-T_{\zeta})/\Delta T_\zeta\right)+1\right)$, where $T_{\zeta}=160$MeV, $\Delta T_\zeta=4$MeV.
The equation of state is an interpolation of lattice QCD results at high temperatures
\cite{Borsanyi:2010cj} and a hadron gas model equation of state at lower temperatures. 
In constructing the equation of state we follow the procedure of
 \cite{Chojnacki:2007jc}. The temperature dependence 
of the sound velocity has no soft point \cite{Bozek:2011ua}.

The hydrodynamic evolution stops at the freeze-out temperature of $135$MeV.
At the freeze-out hypersurface particle emission is done following the Cooper-Frye 
formula in the  event generator THERMINATOR \cite{Chojnacki:2011hb}, 
 with viscous corrections to the equilibrium  momentum distribution $f_0$
\begin{equation}
f=f_0+\delta f_{shear}+\delta f_{bulk}
\label{eq:deltaf}
\end{equation}
We use quadratic corrections in momentum  for the shear viscosity 
\begin{equation}
\delta f_{shear}= f_0
\left(1\pm f_0 \right) \frac{1}{2 T^2 (\epsilon+p)}p^\mu p^\nu \pi_{\mu\nu}
\label{eq:dfsh}
\end{equation}
and asymptotically linear corrections  for the bulk viscosity
\begin{equation}
\delta f_{bulk}= C_{bulk}f_0
\left(1\pm f_0 \right)\left(c_s^2 u^\mu p_\mu -\frac{(u^\mu p_\mu)^2-m^2}
{3 u^\mu p_\mu}\right) \Pi  \ ,
\end{equation}
with 
\begin{equation}
\frac{1}{C_{bulk}}= \frac{1}{3}\sum_n\int \frac{d^3 p}{(2\pi)^3}\frac{m^2}{E}f_0
\left(1\pm f_0 \right)\left(c_s^2 E -\frac{p^2}{3 E}\right) \ ,
\label{eq:dfbu}
\end{equation} 
 the sum runs over all the hadron species.

\begin{figure}
\includegraphics[width=.3\textwidth]{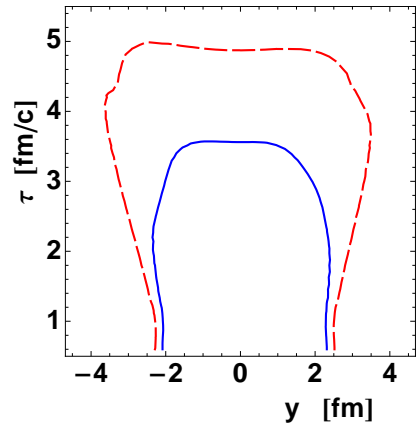}
\caption{(Color online) Constant temperature
 hypersurface $T(\tau,x=0,y,\eta_{\parallel}=0)$ in a p-Pb 
interaction for the freeze-out temperature 
$T_f=135$MeV (dashed line) and for $160$MeV (solid line). }
\label{fig:freezep}
\end{figure}

In Fig. \ref{fig:freezep} is shown the freeze-out hypersurface at 
$\eta_\parallel=0$ for a p-Pb event with $N_{part}=24$. The dense source 
survives for $5$fm/c with the lifetime of the deconfined phase  of $3.5$fm/c 
($T>160$ MeV, solid line contour in Fig. \ref{fig:freezep}).

\section{Results}

\begin{figure}
\includegraphics[width=.48\textwidth]{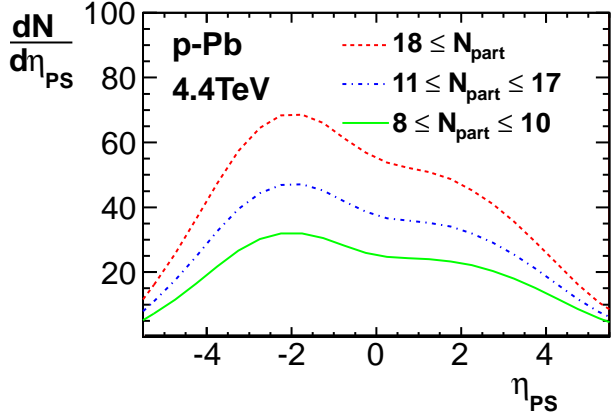}
\caption{(Color online) Pseudorapidity distribution of charged particles in p-Pb interactions at $\sqrt{s_{NN}}=4.4$TeV. The dashed, dashed-dotted and solid lines correspond to  the three centrality 
classes defined by  the number of participant nucleons. 
The distributions are  shown in the laboratory frame for the LHC experiments. }
\label{fig:peta}
\end{figure}

\begin{figure}
\includegraphics[width=.48\textwidth]{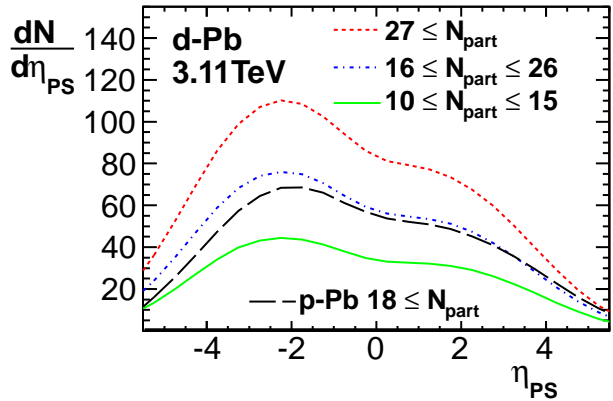}
\caption{(Color online) Same as Fig. \ref{fig:peta} but for d-Pb interactions 
at $\sqrt{s_{NN}}=3.11$TeV. 
 The long-dashed
 line shows the pseudorapidity density for the most central p-Pb collisions at $4.4$TeV.}
\label{fig:deta}
\end{figure}

\begin{figure}
\includegraphics[width=.38\textwidth]{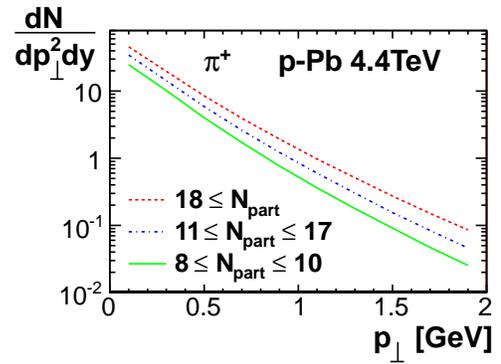}
\caption{(Color online) Transverse momentum spectra of $\pi^+$ in p-Pb interactions
 at $\sqrt{s_{NN}}=4.4$TeV, for $y=0$ in the laboratory frame.
 The dashed, dashed-dotted and solid lines correspond 
to the three centrality 
classes defined by  the number of participant nucleons. }
\label{fig:ppion}
\end{figure}

\begin{figure}
\includegraphics[width=.38\textwidth]{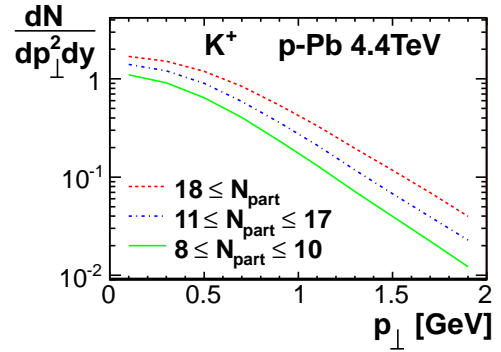}
\caption{(Color online) Same as Fig. \ref{fig:ppion} but for  $K^+$. }
\label{fig:pkaon}
\end{figure}

For each centrality class $50$ hydrodynamic event are calculated.
 For each event several hundred THERMINATOR events are generated and analyzed
 together. This reduces non-flow effects, which in this case come from 
resonances decays. The numerical gird for the hydrodynamic evolution
is set in the NN c.m. frame. The momenta of the emitted particles are
boosted by $y_{sh}=0.46$ and $0.12$ for p-Pb and d-Pb collisions to obtain
spectra in the LHC laboratory frame.

\begin{figure}
\includegraphics[width=.38\textwidth]{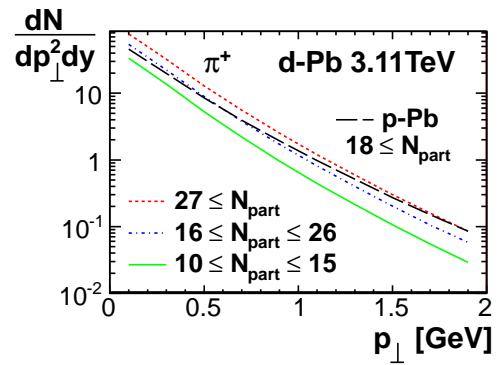}
\caption{(Color online) Same as Fig. \ref{fig:ppion} but for d-Pb interactions at $\sqrt{s_{NN}}=3.11$TeV. }
\label{fig:dpion}
\end{figure}

\begin{figure}
\includegraphics[width=.38\textwidth]{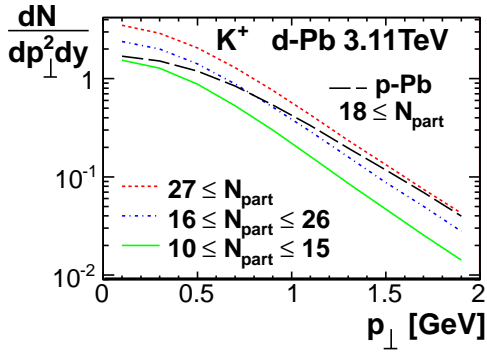}
\caption{(Color online) Same as Fig. \ref{fig:dpion} but for $K^+$. }
\label{fig:dkaon}
\end{figure}

The distribution of charged particles in pseudorapidity 
is shown in Fig. \ref{fig:peta} for the three centrality classes
 defined in Sec. \ref{sec:ini}. The density of 
charged particles at midrapidity for centrality bins $N_{part}\ge 18$ and $11\le N_{part}\le 17$ in
p-Pb is larger than observed in Pb-Pb collisions at $2.76$TeV for 
centrality $70-80$\%. One can expect a similar degree of collective 
acceleration as in peripheral Pb-Pb events. The particle multiplicity in p-Pb
interactions is of the same order as in p-p interactions with the 
highest multiplicity
analyzed by the CMS collaboration \cite{Khachatryan:2010gv}. While the nature
of the high multiplicity p-p 
events is still unclear, the multiplicity in a p-Pb
 or d-Pb collision is simply related to the source size and density.

The charged particle densities in pseudorapidity for p-Pb (Fig. \ref{fig:peta})
and d-Pb collisions (Fig. \ref{fig:deta}) are asymmetric, reflecting 
the predominant emission from the participant nucleons in the Pb nucleus 
\cite{Bialas:2004su}. For d-Pb  collisions in the centrality bin $N_{part}\ge 27$ the 
particle multiplicity is similar as in $60-70$\% centrality Pb-Pb 
interactions. This makes the applicability of the hydrodynamic model even 
more justified in that case. The particle multiplicity in p-Pb events with  $N_{part}\ge 18$
  is similar as for d-Pb events with $16\le N_{part}\le 26$. We find, that in all the cases
the charged particle density at midrapidity is to within $5$\%
proportional to the  number of participant nucleons.

The transverse momentum spectra for $\pi^+$ and $K^+$ emitted in p-Pb collisions 
are hardened by the collective transverse flow generated in the hydrodynamic
 expansion (Figs. \ref{fig:ppion} and \ref{fig:pkaon}). For more central 
collisions  the spectra are slightly flatter, as more 
transverse flow is generated. A similar picture appears for transverse momentum
spectra in d-Pb collisions. The spectra from  $N_{part}\ge 27$  and  $16\le N_{part}\le 26$ centrality bins
have a similar effective slope and are harder than the ones for the $10\le N_{part}\le 15$  
centrality class. It is interesting to observe that the transverse momentum 
spectra are harder in p-Pb than in d-Pb collisions. The transverse size of 
the fireball in p-Pb interactions is smaller but its density is higher, this
leads to a faster transverse expansion than for d-Pb interactions.

\begin{figure}
\includegraphics[width=.35\textwidth]{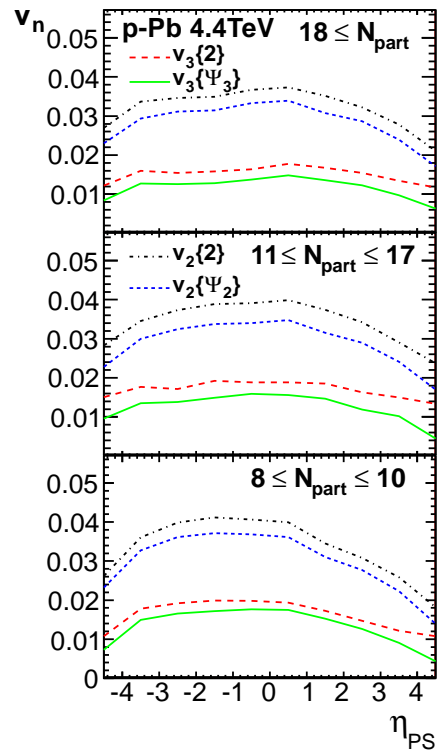}
\caption{(Color online) The elliptic and triangular flow coefficients
 of charged particles as function of pseudorapidity in the laboratory frame in
 p-Pb interactions, for centralities
 $0-4$\% (top panel), $4-32$\% (middle panel),
and $32-49$\% (bottom panel).
The dashed-dotted and dotted lines represent the 
elliptic flow coefficients $v_2\{2\}$ and $v_2\{\Psi_{2}\}$,  while the dashed and solid lines
represent the triangular flow coefficients  $v_3\{2\}$ and $v_3\{\Psi_{3}\}$.}
\label{fig:pv2eta}
\end{figure}

\begin{figure}
\includegraphics[width=.35\textwidth]{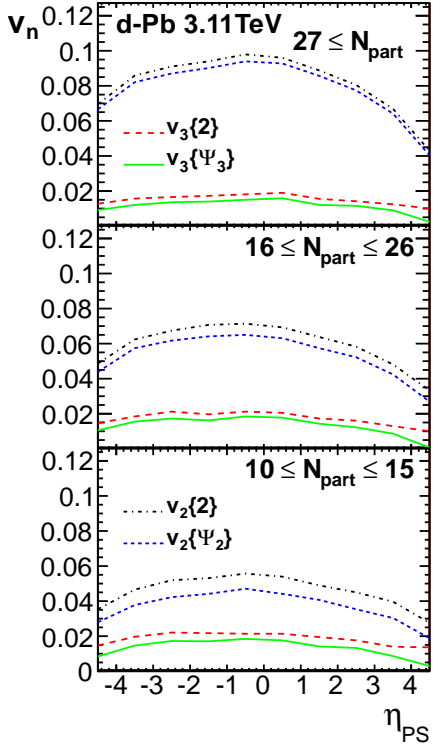}
\caption{(Color online) Same as Fig. \ref{fig:pv2eta} but for
 d-Pb interactions, for centralities
 $0-5$\% (top panel), $5-30$\% (middle panel),
and $30-50$\% (bottom panel).}
\label{fig:dv2eta}
\end{figure}

\begin{figure}
\includegraphics[width=.38\textwidth]{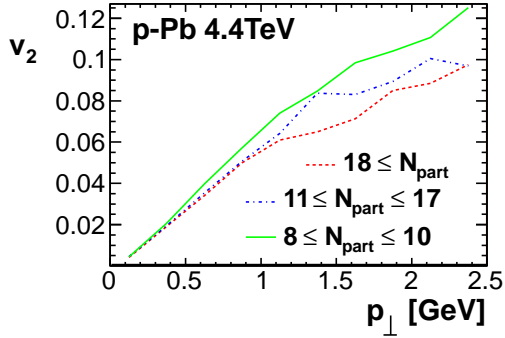}
\caption{(Color online) The elliptic flow coefficient of charged particles 
as function of transverse momentum around
$y=0$ in the laboratory frame for p-Pb interactions.  The dashed, dashed-dotted and solid lines correspond 
to the three centrality 
classes defined by  the number of participant nucleons, $N_{part}\ge 18$, $17\ge 
N_{part}\ge 11$ and $10\ge N_{part}\ge 8$ respectively.}
\label{fig:v2ptp}
\end{figure}

\begin{figure}
\includegraphics[width=.38\textwidth]{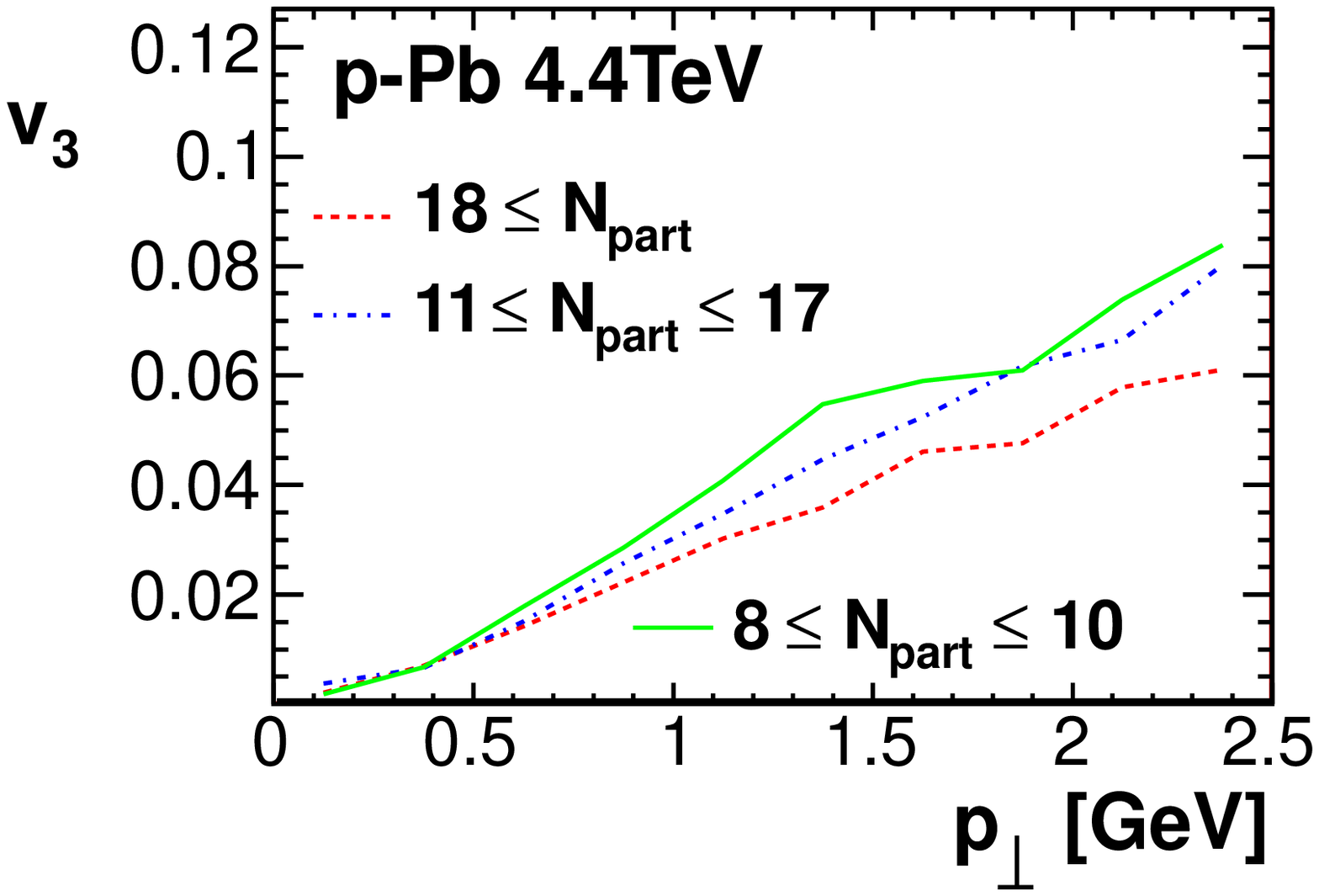}
\caption{(Color online) Same as Fig. \ref{fig:v2ptp} but for the triangular flow.}
\label{fig:v3ptp}
\end{figure}

\begin{figure}
\includegraphics[width=.38\textwidth]{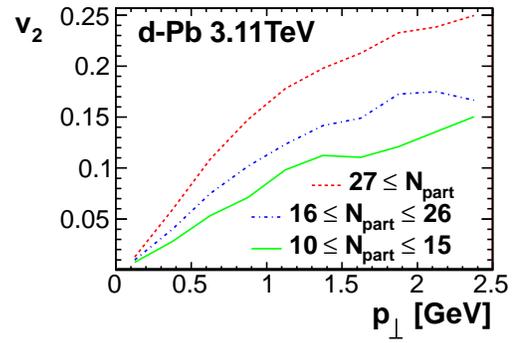}
\caption{(Color online) Same as Fig. \ref{fig:v2ptp} but d-Pb interactions.
The dashed, dashed-dotted and solid lines correspond 
to the three centrality 
classes defined by  the number of participant nucleons, $N_{part}\ge 27$, $26\ge 
N_{part}\ge 16$ and $15\ge N_{part}\ge 10$ respectively. }
\label{fig:v2ptd}
\end{figure}

\begin{figure}
\includegraphics[width=.38\textwidth]{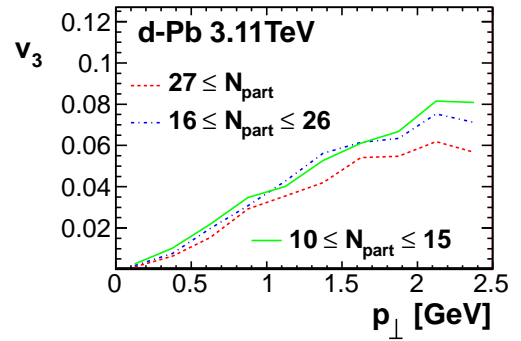}
\caption{(Color online) Same as Fig. \ref{fig:v2ptd} but for the triangular flow.}
\label{fig:v3ptd}
\end{figure}

Fluctuating initial densities
 (Fig. \ref{fig:inidxy}) have a nonzero eccentricity 
and triangularity (Fig. \ref{fig:e2pPb} and \ref{fig:e2dPb}). The short 
hydrodynamical expansion stage in these  systems is sufficient 
to generate  noticeable elliptic and triangular flows.
Fig. \ref{fig:pv2eta} shows the pseudorapidity dependence of the
  $p_\perp$ integrated elliptic  $v_2$ and triangular $v_3$ flow coefficients in p-Pb interactions. 
In the calculations, $500$ to $1500$ THERMINATOR  events are generated from each hypersurface generated 
in a $3+1$-D viscous hydrodynamic evolution. The event plane orientations $\Psi_2$ and $\Psi_3$ are found
and the elliptic $v_2$ and triangular $v_3$ flow coefficients are calculated in each event. 
The average over the hydrodynamic events gives $v_2\{\Psi_2\}=\langle v_2 \rangle$ and $v_3\{\Psi_3\}
=\langle v_3 \rangle$. The second cumulant flow coefficients include flow fluctuations 
$v_n\{2\}=\sqrt{\langle v_n^2 \rangle}$.
A moderate dependence of the elliptic and triangular flows
 on centrality is seen.
In p-Pb collisions both the eccentricity and the triangularity deformations
 of the initial shape
are fluctuation dominated. We observe some reduction of the collective flow at 
forward and backward pseudorapidities. This reduction
 is due to an increase of dissipative effects and a shorter life-time of 
the source at nonzero space-time rapidities \cite{Hirano:2005xf,Bozek:2009mz}.
The form of the pseudorapidity dependence of the harmonic coefficients of
 the flow in Fig. \ref{fig:pv2eta} must be taken with caution, as 
 $3+1$-D viscous hydrodynamic calculations cannot reproduce it
accurately \cite{Schenke:2010rr,Bozek:2011ua}.
The azimuthally asymmetric flow is different in d-Pb collisions 
(Fig. \ref{fig:dv2eta}). The elliptic flow is significantly larger than
 in p-Pb interactions, reaching $0.097$ for central collisions. 
We notice a strong centrality dependence, $v_2$ increases significantly 
for central collisions. The initial eccentricity of the 
source in d-Pb collisions is large (Fig. \ref{fig:e2dPb}). 
The elliptic flow fluctuations 
(the difference between $v_2\{2\}$ and $v_2\{\Psi_{2}\}$ \cite{Voloshin:2007pc}) 
are relatively less important for the deuteron than for the proton induced interactions.
The triangular flow in d-Pb is similar as in p-Pb  collisions, and does not
vary strongly with the centrality.

The hydrodynamic response translates the initial deformation of 
the fireball into the  azimuthal 
asymmetry of the final flow in an event by event basis. The  elliptic flow coefficient 
follows closer the initial eccentricity than the triangular flow 
follows  the initial triangularity, both in the 
magnitude and the orientation of the event plane. This observation is 
in agreement with other studies
\cite{Gardim:2011xv,*Chaudhuri:2011pa}. The hydrodynamic response $v_2/\epsilon_2$ 
is larger for central collisions, where dissipative effects that 
reduce the flow asymmetry are smaller \cite{Drescher:2007cd}.


The elliptic and triangular flow coefficients show a hydrodynamic behavior 
as function of the transverse momentum (Figs. \ref{fig:v2ptp} and 
\ref{fig:v2ptd}).  Same as for the integrated flow, there is little 
change with 
centrality for the elliptic flow, while some decrease of $v_3$
 in most 
central p-Pb collisions is seen. The elliptic flow $v_2(p_\perp)$ 
for the d-Pb system (Fig. \ref{fig:v2ptd}) is
large, it increases significantly for central collisions, where a 
saturation of the dependence on the transverse momentum appears around $1$GeV.
The triangular flow in d-Pb interactions (Fig. \ref{fig:v3ptd}) is similar in 
magnitude as in p-Pb collisions, and shows almost no variation with the
 centrality.

\section{Conclusions}

The formation of a hot, collectively expanding fireball in p-Pb collisions at
$\sqrt{s_{NN}}=4.4$TeV and d-Pb collisions at $3.11$TeV is studied.
We perform $3+1$-D event by event viscous hydrodynamic calculations.
The initial size and shape of  the fireball is taken from the
 Glauber Monte-Carlo model. The initial entropy density is adjusted to 
reproduce
the expected particle multiplicity estimated as an extrapolation from 
observations in  
peripheral Pb-Pb collisions at $\sqrt{s_{NN}}=2.76$TeV.

\begin{figure}
\includegraphics[width=.38\textwidth]{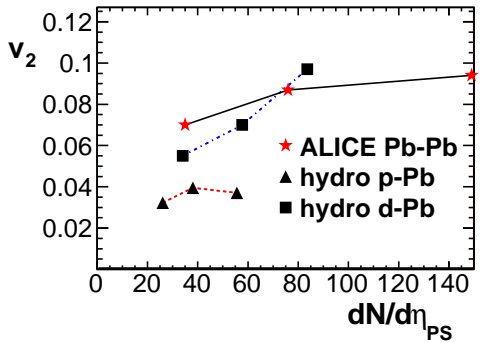}
\caption{(Color online) Elliptic flow coefficient $v_2\{2\}$ as 
function 
charge particle density at central pseudorapidity. Hydrodynamic 
calculations for p-Pb collisions at $\sqrt{s_{NN}}=4.4$TeV (triangles), 
for d-Pb collisions at $3.11$TeV (squares), 
and experimental data from the ALICE collaboration for 
Pb-Pb collisions at $2.76$TeV (stars) \cite{Aamodt:2011vk} are shown.}
\label{fig:v2all}
\end{figure}

A small hot and dense fireball is formed. It expands rapidly in the 
transverse direction. The $p_\perp$ spectra of emitted particles get harder, 
especially for p-Pb collisions. The deconfined phase survives for $3-4$fm/c
in events with high particle multiplicity, the presence of such a dense medium 
should be visible in the 
  nuclear attenuation  factor for high $p_\perp$ hadrons.
The size and the 
life-time of the source can be further constrained in same-pion 
interferometry measurements.
 The initial eccentricity and triangularity of a
lumpy initial fireball lead to the formation of an azimuthally 
asymmetric flow. The elliptic flow is $3-4$\% in p-Pb collisions, with little
 centrality dependence (Fig. \ref{fig:v2all}).  For the d-Pb system, the 
elliptic flow is significantly larger, increasing for central collisions, 
and reaching  almost $10$\%. A comparison to peripheral Pb-Pb collisions 
in Fig. \ref{fig:v2all} shows that similar conditions are realized in 
proton or deuteron induced interactions. The elliptic flow of that magnitude 
can be measured, with a different dependence on centrality in p-Pb and d-Pb 
collisions.

Let us close with a discussion on future prospects for proton and deuteron 
induced reactions at ultrarelativistic energies. p-Pb collisions at $4.4$TeV
are planned in the near future at the LHC \cite{Salgado:2011wc}. The 
elliptic flow and the hardening of the $p_\perp$ spectra are 
noticeable and should be
looked for in the experimental analysis. However, it must be stressed that 
the dynamics of such small systems is at the 
limit of the applicability of the viscous hydrodynamic model. The use of the 
hydrodynamic model in d-Pb interactions is better justified, also the elliptic
 flow is stronger, 
but such collisions are  not planned in the near future at the LHC.
The shift to the maximum LHC energy to $\sqrt{s_{NN}}=6.22$ for d-Pb and $8.8$TeV
for p-Pb collisions results in an increase of the particle multiplicity by 
$30$\%. Hydrodynamic expansion  would last longer, with less dissipative 
effects. The eccentricity and triangularity 
are similar as at 
lower LHC energies and qualitatively we expect similar results.

In view of the results in this paper it seems very interesting to 
look for collective effects in d-Au collisions at $\sqrt{s_{N}}=200$GeV
in  RHIC experiments.
The multiplicity in central d-Au interactions is similar as in peripheral 
Au-Au collisions at the same energy. If some stage of 
collective expansion is present, the large initial 
eccentricity in a d-Au system should translate into a
measurable elliptic flow. Unfortunately no published data exist for
 these experiments, hydrodynamical simulations are underway.

\section*{Acknowledgments}
Supported by 
Polish Ministry of Science and Higher Education under
grant N N202 263438. Numerical calculations were made on the Cracow Cloud 
One cluster.
\bibliography{../hydr}


\end{document}